\documentclass[article,showpacs,twocolumn,superscriptaddress,amssymb]{revtex4-1}
\usepackage{graphicx}
\usepackage{bm}
\usepackage{xcolor}
\usepackage{epsfig}
\usepackage{amsfonts}
\usepackage{mathtools}
\usepackage{amsmath}
\newcommand{\beqa}{\begin{eqnarray}}
\newcommand{\beeq}{\begin{equation}}
\newcommand{\eeqa}{\end{eqnarray}}
\newcommand{\eeqe}{\end{equation}}

\begin{document}

\title{Derivation of the spin Hamiltonians for  Fe in MgO}

\author{A. Ferr\'on}
\affiliation{International Iberian Nanotechnology Laboratory (INL), Av. Mestre Jos\'e Veiga, 4715-330, Braga, Portugal}

\affiliation{Instituto de Modelado e
Innovaci\'on Tecnol\'ogica (CONICET-UNNE), Avenida Libertad 5400, W3404AAS
Corrientes, Argentina.}

\author{F. Delgado}

\author{J. Fern\'andez-Rossier}

\altaffiliation{Permanent address: Departamento de F\'{\i}sica Aplicada, 
Universidad de Alicante.}

\affiliation{International Iberian Nanotechnology Laboratory (INL), Av. Mestre Jos\'e Veiga, 4715-330, Braga, Portugal}

\begin{abstract}
A method to calculate the effective spin Hamiltonian for a transition metal impurity in a non-magnetic insulating host is presented and applied to the paradigmatic case of Fe in MgO. In a first step we calculate the electronic structure
employing    standard  density functional  theory (DFT), based on generalized-gradient approximation (GGA),  using plane waves as a basis set.
 The corresponding basis of atomic-like 
  maximally localized Wannier 
functions is derived and used to represent the DFT Hamiltonian,
resulting in a tight-binding model for the atomic orbitals of the magnetic 
impurity.  The third step is to solve, by exact numerical 
diagonalization, the $N$ electron problem in the open shell of the magnetic 
atom,
including  both  effect of spin-orbit and Coulomb repulsion. Finally, the low energy sector of this multi-electron Hamiltonian is 
mapped into effective spin models that, in addition to the spin matrices $S$, 
can also include the orbital angular momentum $L$ when appropriate. We successfully apply the method to
 Fe in MgO,  considering both,  the undistorted and Jahn-Teller (JT) 
distorted cases.  Implications for the influence of Fe impurities 
on the performance of magnetic tunnel junctions based on MgO are discussed. 

\end{abstract}
\date{\today}

\pacs{73.22.-f,73.22.Dj}
\maketitle

\section{Introduction}
Understanding the electronic properties of magnetic transition metals embedded 
in diamagnetic hosts plays a central role in several branches of condensed 
matter physics and materials science. The presence of transition metal 
impurities is known to modify the electronic properties of 
insulators \cite{Abragam}, semiconductors \cite{Furdyna} and molecular crystals \cite{Sessoli}. 
Thus,  diluted  semiconductors become paramagnetic and their optoelectronic 
properties, such as the photoluminescence spectrum become extremely sensitive 
to the application of   magnetic fields, resulting in the so called giant 
Zeeman splitting \cite{Furdyna}. In turn, the electronic and spin  properties 
of the magnetic atoms are very sensitive to their environment \cite{Abragam}.
This permits inferring  local information about the host by means of spin probing
techniques such as electron paramagnetic resonance \cite{Abragam}.

Very often, the spin properties of a magnetic system are described in terms of 
effective single spin Hamiltonians \cite{Abragam,Sessoli} built in terms of 
atomic spin operators only. Whereas the symmetry of a given system determines 
which terms are possible in an effective spin Hamiltonian,  prediction of the 
values of the various parameters can be a difficult problem.  Extraordinary 
progress in instrumentation techniques  makes it now possible to probe 
individual magnetic atoms in a solid state 
environment \cite{solotronics1,solotronics2} using a variety of techniques, 
such as scanning tunneling microscope (STM) inelastic electron spectroscopy 
(IETS) \cite{Hirjibehedin2007,Heinrich2014},  and single quantum dot photoluminescence \cite{Besombes2004,Leger2006}. These techniques permit  assessing the delicate 
interplay between spin properties of the transition metal and electronic and 
structural properties of the local environment at the atomic scale \cite{Oberg14,Heinrich2014} 
and motivate the quest of quantitative methods to address this interplay.

Conventional density functional theory \cite{Kohn,Kohn-Sham} provides an accurate description of the  
electronic properties of the ground state of  solids but it does not provide a direct route  to describe 
the fine details of 
the low energy spin excitations inherent in magnetic atoms in insulating hosts.
For instance,  the ground state of the effective Hamiltonian in conventional 
functionals in DFT is a unique Slater determinant with broken spin symmetry, 
which is fundamentally different from the multiplet  nature of the real system.
In this context, we find it  convenient to have a constructive theoretical 
approach to derive the effective spin Hamiltonian, starting from an atomistic 
DFT description of the electronic properties of the system, but describing the 
electronic properties of the system with a multi-electron approach that 
captures the multiplet nature of the relevant electronic states.  

Here we propose a method to  obtain an effective spin Hamiltonian for a 
magnetic atom in an insulating host,  starting from density functional 
calculations, in four well defined steps. First, a density functional 
calculation of the electronic properties of the magnetic atom inside the 
non-magnetic host.  The second step is to represent the effective DFT 
Hamiltonian with a basis of localized atomic orbitals, which allows obtaining 
the crystal and ligand fields terms of  the atomic orbitals of the relevant 
open shell of the magnetic atom, defining thereby a multi-orbital Hubbard Hamiltonian.
 Since our DFT approach makes use of a  
plane-wave basis, we implement this step by means of the 
wannierization \cite{wan1} technique.
Up to this point, the methodology is very similar to previous work \cite{Miyake2006,lorente-wan1,lorente-wan2,Haverkort12,dpmodel,cf-wan,refe1,refe2}.   
In a third step we add to the Hubbard model the intra-atomic Coulomb repulsion  and the spin-orbit coupling
 for the electrons in the open-shell.  The final step is a 
symmetry analysis of the spectrum and wave functions, obtained by numerical diagonalization of the 
effective Hubbard model. The resulting 
multi-electron states analysis permits  constructing an effective spin Hamiltonian 
for the system.

Below we describe in more detail the method and  apply it to the paradigmatic
case of Fe$^{2+}$ as a substitutional impurity of Mg in MgO \cite{Abragam}, a band insulator.
The spin properties of this system have been studied in detail by means of several techniques, including
 far infrared spectroscopy \cite{Wong68},
 acoustic paramagnetic resonance \cite{APR72}, infrared spectroscopy \cite{Ham77}, and XPS \cite{Haupricht2010}.
 The interplay between atomic structure and 
spin properties is beautifully illustrated in this system: we consider both 
the case of undistorted Fe/MgO, where the octahedral symmetry does not quench 
completely the orbital angular momentum $L$  of Fe$^{2+}$ as well as the 
system with a Jahn Teller distortion in which case $L$ is quenched, resulting 
in a very different type of effective spin Hamiltonian.  Our findings might 
shed some light on recent results \cite{Heinrich2014}  probing a single Cobalt 
atom on MgO, that indicate that the orbital moment is  partially quenched, 
in contrast with magnetic adatoms deposited on other surfaces. In addition, we are interested in Fe 
as a possible impurity in MgO tunnel barriers in magnetic tunnel junctions with Fe based electrodes \cite{yua04,par04} and 
we discuss how it could reduce the spin-filter properties, when compared to 
the ideal system. 

The rest of this manuscript is organized as follows. In Sec. \ref{sdft}
we study the Electronic Structure of Fe$^{2+}$ as a substitutional 
impurity of Mg in MgO using DFT calculations. In Sec. \ref{swan} we
discuss the derivation of the single-particle part of the magnetic atom Hamiltonian
 from the DFT calculation using
the wannierization approach. In Sec. \ref{CI}, we build and solve by numerical diagonalization
the generalized Hubbard model and derive the
effective spin Hamiltonians for two different geometries. In Sec. \ref{con} we summarize the advantages and shortcomings of this
work and discuss the effect of Fe impurities in MgO barriers on the magnetoresistance of  magnetic tunnel junctions.

\section{Electronic Structure: DFT Calculations \label{sdft}}

In this section we describe our DFT calculations, for  pristine MgO as well  
calculations for super cells of Mg$_{31}$O$_{32}$Fe.  For the super cells,  
we consider two geometries, with and without Jahn-Teller distortion of the Fe 
atom. In addition, and for reasons discussed below, we did both spin-polarized 
and spin-unpolarized calculations. 

Our calculations were done 
 using the generalized-gradient approximation (GGA) 
for exchange-correlation energy \cite{gga}, using   plane-wave basis sets and 
ultrasoft pseudopotential method for Mg and O, and Projector Augmented-Wave 
(PAW) \cite{paw} for Fe as implemented in QUANTUM ESPRESSO (QE) code \cite{qe}.
Since we are interested in the spin-unpolarized calculation,  there is no 
need to include the DFT+U correction. Although a proper DFT calculation of 
magnetism will require DFT+U corrections\cite{mgo1}, here we do not include 
the U-term at this level since, as discussed in Secs. \ref{swan} and \ref{CI}, 
our approach to derive an effective spin Hamiltonian requires to start with 
a spin unpolarized DFT calculation to which we should add the many body Coulomb repulsion between the d-orbital electrons in the Fe.

\begin{figure}[h!]
\begin{center}
\includegraphics[width=1.\linewidth]{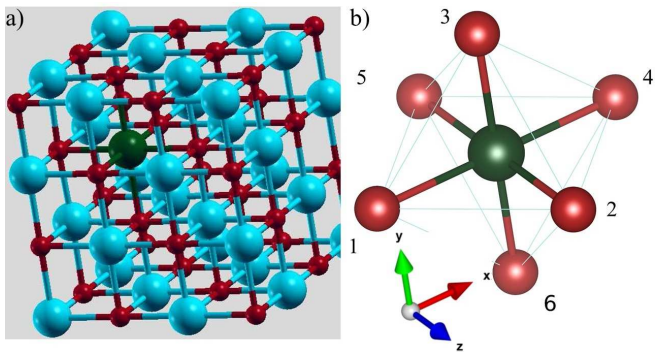}
\end{center}
\vspace{-15pt}
\caption{(Color online) Left panel:  geometric structure for the 
Mg$_{31}$O$_{32}$Fe unit cell used in the DFT calculation:
 Mg atoms in blue, O atoms in red, Fe atom in green.
Right panel shows the octahedral environment, with  Fe surrounded by 6 O atoms.
 }
\label{fst}
\end{figure}

For the case of the super cells, the number of $k$ points was taken to be 
$8\times8\times8$ and we used a Fermi-Dirac smearing with a broadening 
parameter of $0.0035\,Ry$. Finally we fixed the cutoff energies for 
the wave function and charge density at $65\,Ry$ and $700\,Ry$ respectively.
The calculation of the magnetic atom in MgO is done using a $64$ atoms 
supercell of Mg$_{31}$O$_{32}$Fe, with  lattice parameter  $2a$. The supercell,
including the Fe atom is shown in Fig. \ref{fst}(a).

MgO is an insulator with a NaCl type crystal lattice (see Fig. 1).  Using the 
experimental lattice constant ($a= 4.22\,\AA$), our DFT calculations give a 
band gap of  $5.4\, eV$,  below  the actual value  $7.8\, eV$  
 \cite{mgoe1,mgoe2,mgoe3}. This discrepancy is very common and quite close to 
other DFT calculations for MgO (see for example Ref. \cite{mgo1} 
where the calculated gap is $5.85\, eV$).  Our calculations show that
the valence band of the  MgO is mainly formed by O $2p$ orbitals and
the conduction band by Mg $3p$ and $2s$ orbitals.

\begin{figure}[h!]
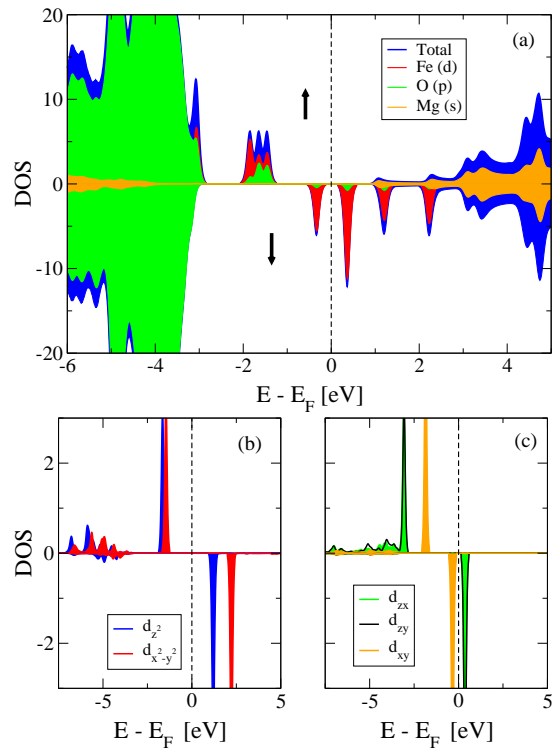

\begin{center}
\includegraphics[width=0.4\textwidth]{fig2a.eps}
\includegraphics[width=0.4\textwidth]{fig2b.eps}
\end{center}
\vspace{-15pt}
\caption{Projected density of states for the ground state of
the distorted Mg$_{31}$O$_{32}$Fe, computed with spin-polarized DFT.
(a) Blue curve:  total density of states (DOS). Red
curve:  DOS projected over $d-$orbitals of Fe. Green and orange: DOS projected over  O $p-$orbitals 
Mg  $s-$orbitals, respectively. Positive values correspond to majority-spin and negative values correspond to minority-spin. 
(b) DOS projected over  $e_g$ orbitals. (c)  DOS projected over 
 the $t_{2g}$ orbitals.
}
\label{fdosp}
\end{figure}

  Our calculations show that the main effect of the Fe impurity on the MgO 
band structure is the appearance of 10 in-gap very narrow bands that, as we 
show below, are associated to the $d$ orbitals of the Fe atom, see Fig. \ref{fdosp}(a). 
Six of these levels are below the Fermi energy and, for spin-polarized 
calculations,  correspond to  5 levels spin $\uparrow$ and 1 $\downarrow$ 
resulting in a spin  $S=2$.

In the ideal MgO-like bulk crystal, where the Fe substitutes a Mg atom, the Fe is in an 
octahedral environment surrounded by 6 oxygen neighbors. In this undistorted geometry, the Fe-$d$ 
levels are expected to split in a lower energy triplet, $t_{2g}$, and a higher 
energy doublet $e_g$, due both to the interaction with the charged  neighbor 
oxygens (crystal field contribution) and the hybridization with the oxygen 
atomic orbitals (ligand field contributions).

In the undistorted octahedral environment,  the Fermi energy lies exactly at the $t_{2g}$ 
orbital triplet of the minority spin, so that the system has a 
orbital degeneracy that leads  to Jahn-Teller instability \cite{Abragam,Ham77,Haupricht2010},
 which we model by letting the system relax 
from an initial configuration in which Fe is slightly off the center of the octahedron. The distorted solution so found has 
 lower energy than the undistorted one.  In both cases, distorted and 
undistorted,
the relaxation was performed until the forces acting on atoms were smaller 
than $10^{-3}\,a.u.$.
In the undistorted octahedral environment, the O surrounding atoms are all 
at $2.135 \AA$ from the Fe atom.  In order to characterize the 
deformed configuration it is convenient to set Fe as the 
origin of coordinates and label oxygens as in the right panel of Fig. \ref{fst}, 
with coordinates $\vec{R}=(X_i,Y_i,Z_i)$ , $i=1,..,6$\cite{Yosidabook}.  Distortions happen 
to be symmetric, {\em i.e.}, with  $\delta \vec{R}_{1}=-\delta \vec{R}_{4}$,  
and we express them in terms of the normal modes of the octahedron.  It turns 
out \cite{Yosidabook} that the computed distortion can be expressed as a linear combination of 
the breathing mode $q_1=\frac{1}{\sqrt{3}}(X_1+Y_2+Z_3)$, which clearly 
preserves the octahedral symmetry, and the 
$q_3=\frac{1}{\sqrt{6}}(2Z_3-X_1-Y_2)$, which singles out  the $z$ axis 
symmetrywise and preserves the planar square symmetry of the $xy$ plane.  
The obtained distortion can be written as $0.01 q_1+ 0.03 q_3$, 
where $q_1$ are expressed in $\AA$ and is said to be tetragonal 
\cite{Yosidabook}.
It should be emphasized that we have not made a systematic attempt to study 
all possible Jahn-Teller distortions in this system. Instead, we are testing 
our method for a particular distortion, which is in line with previous work \cite{Haupricht2010}.

\begin{figure}[h]
\begin{center}
\includegraphics[angle=270,width=1.0\linewidth,clip]{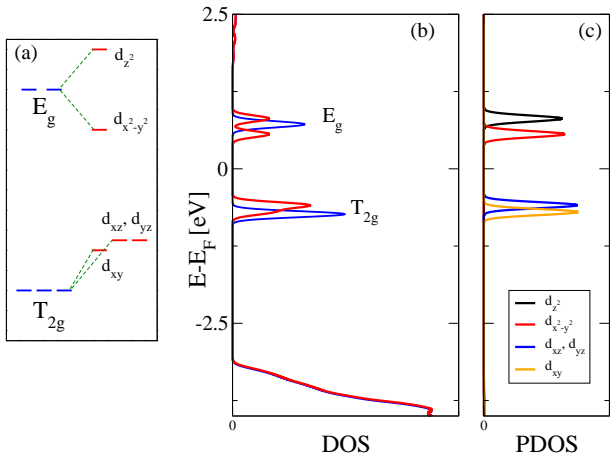}
\end{center}
\caption{(Color online) Spin-unpolarized calculations for the distorted and
undistorted cases. (a) Schematic energy diagram of the $d$ orbitals with
and without Jahn-Teller distortion. (b) Total density of states of
the undistorted case (blue line) and the distorted one (red line).
(c) Projected density of states over different $d$ orbitals for the 
distorted case.}
\label{fdosnm}
\end{figure}

The effect of the tetragonal distortion is apparent in both, the spin-polarized, Fig. \ref{fdosp}, and the spin-unpolarized cases, Fig. \ref{fdosnm}.
The finite width   $\sim 50\, meV$ of the DOS peaks, much 
smaller than the crystal field splitting, is a finite size effect due 
hybridization of $d$ orbitals between Fe atoms located at  different unit cells.
In both cases the $t_{2g}$ triplet degeneracy is split into  a doublet and a 
singlet, and the $e_g$ doublet is also split.
Importantly, the tetragonal distortion does preserve the  $4 \mu_B$  magnetic moment ($S=2$).
However, the very different 
orbital arrangement will result in important differences in the spin Hamiltonian, discussed below.

The spin-polarized calculations discussed so far provide a  mean-field-like 
description of the magnetism of Fe in MgO. However, in order to  determine 
the parameters for the Hamiltonian in the multiplet configuration interaction 
calculation, presented in section IV,  we
 start from a 
spin-unpolarized calculation, a strategy used as well in previous  work \cite{refe2}.  
It must be noted that, for  spin-polarized calculations, the crystal field splitting 
$\Delta$ is spin dependent, which is clearly a feature of a mean field 
solution that breaks spin-rotational symmetry.  In the distorted case the sign 
of the splitting of the $t_{2g}$ levels, as well as the magnitude of the 
splitting of the $e_g$ levels are  spin-dependent.  Since it is convenient to 
have a spin-independent crystal field Hamiltonian, we have performed a 
spin-unpolarized calculation of Fe in MgO.   For the undistorted case we 
obtain a 
ground state configuration $(0e_g,6t_{2g})$ where all $d-$electrons of Fe 
occupy the degenerate (orbital and spin) states $d_{xy}$, $d_{yz}$ and $d_{xz}$ [see Fig. \ref{fdosnm}a),b)].
The computed crystal field splitting is  $\Delta=1.45$ eV. 
For the tetragonal distortion, the spin unpolarized calculation still shows  that 6 electrons occupy the 
$t_{2g}$ levels,  but the $d_{xy}$ level is now split from $d_{xy}$ and 
$d_{yz}$, as shown in Fig. \ref{fdosnm}c). 
 
\section{ Calculation of the crystal field Hamiltonian using Wannier functions \label{swan}}

The discussion of the previous section shows that it is possible to describe 
Fe in terms of 6 electrons occupying the in-gap levels, which are predominantly
formed by Fe $d$ orbitals. 
To do so, 
we would like to extract from the DFT calculation a one-body Hamiltonian 
projected over these $d$ orbitals that includes their interaction with the 
host crystal. However, the DFT Hamiltonian is expressed in terms of Bloch waves
that in our calculations are expressed as linear combinations of 
plane waves.  In order to go to an atomic like description, we compute the so 
called maximally localized Wannier functions (MLWF) 
 \cite{wan1,wan2,wan3,wan4,wan5}  associated to the Bloch states of the DFT 
calculation, using the package Wannier90. The Wannier functions form an 
orthogonal and complete basis set that we use to express the Hamiltonian. 
Interestingly, we find 5 atomic-like MLWF with the same symmetry than the 
real $\ell=2$ spherical harmonics. Therefore,  we  take the representation of 
the DFT Hamiltonian in this subspace, as the {\it crystal field} 
Hamiltonian  $H_{CF}$ (although it also contains ligand field contributions). 

\begin{figure}[h!]
\begin{center}
\includegraphics[width=0.2\textwidth]{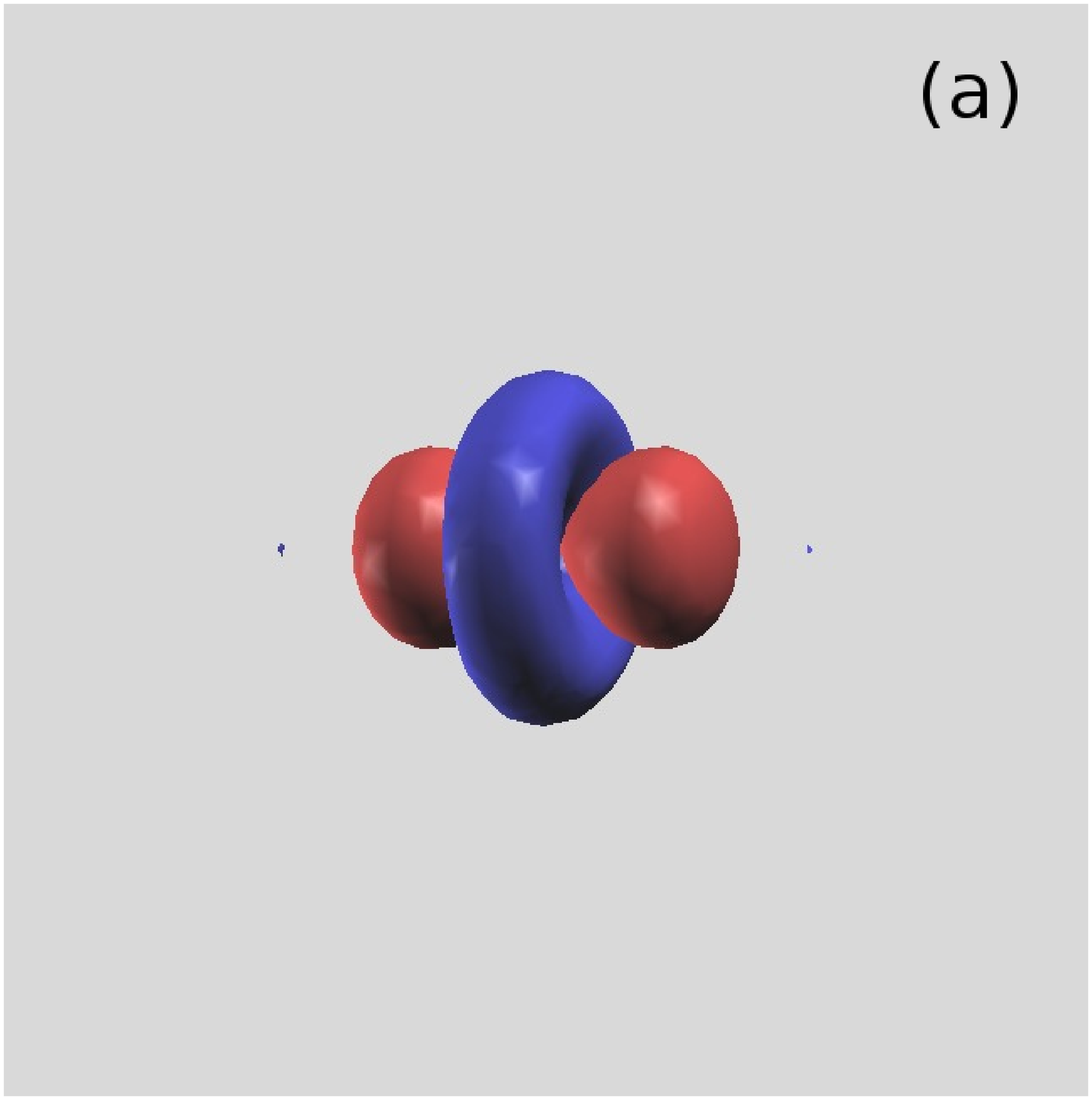}
\includegraphics[width=0.2\textwidth]{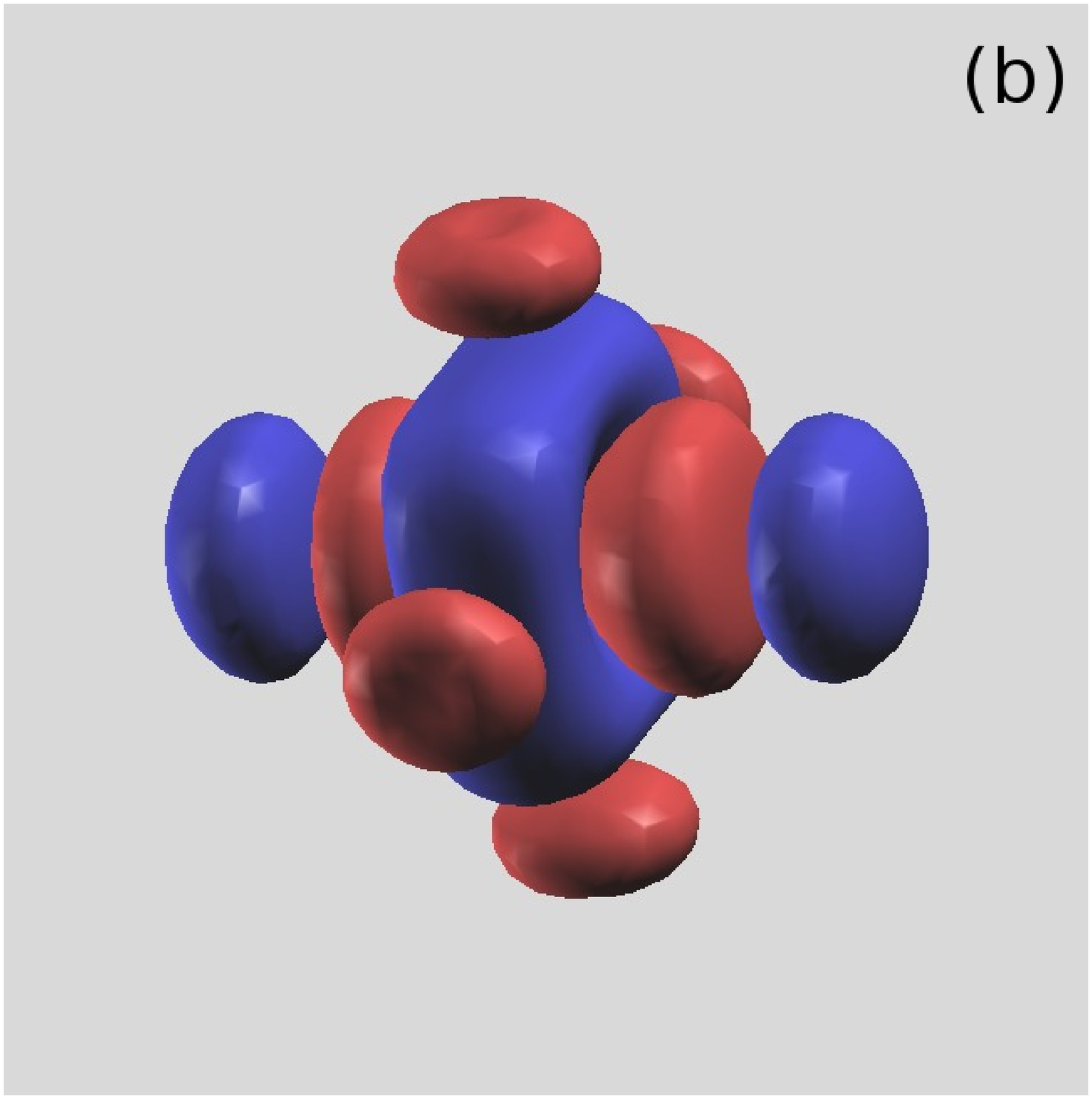}
\includegraphics[width=0.2\textwidth]{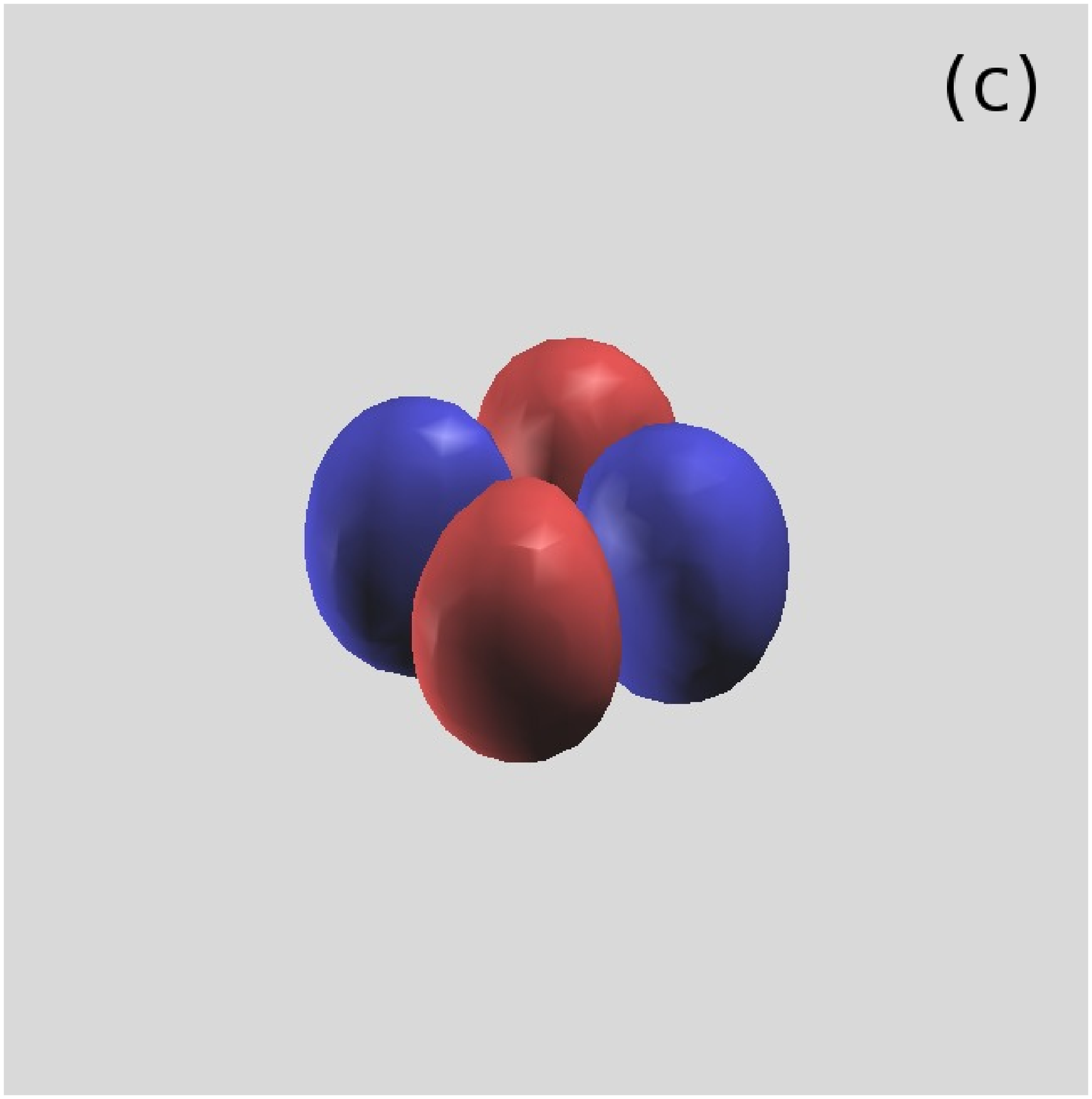}
\includegraphics[width=0.2\textwidth]{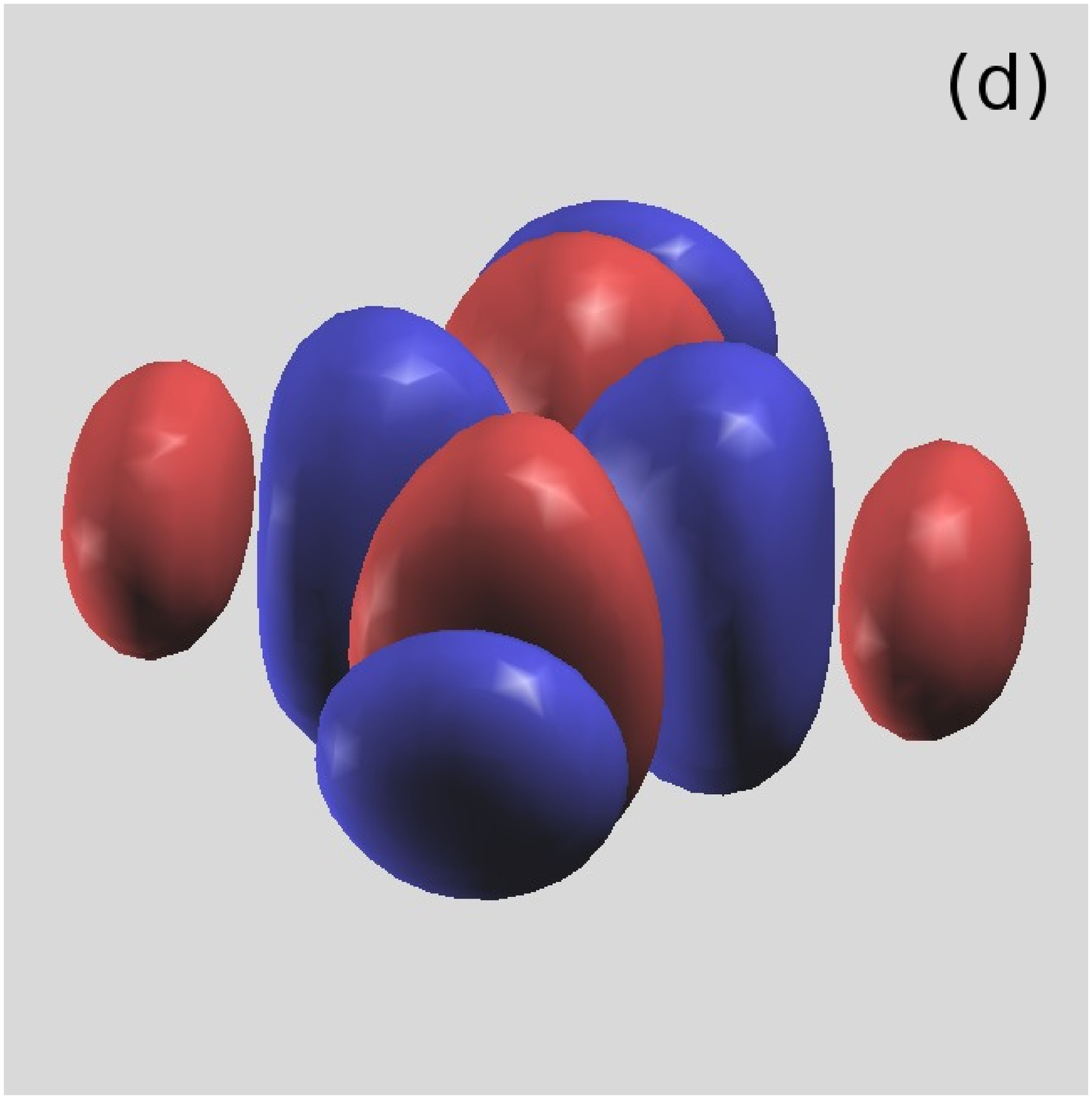}
\includegraphics[width=0.2\textwidth]{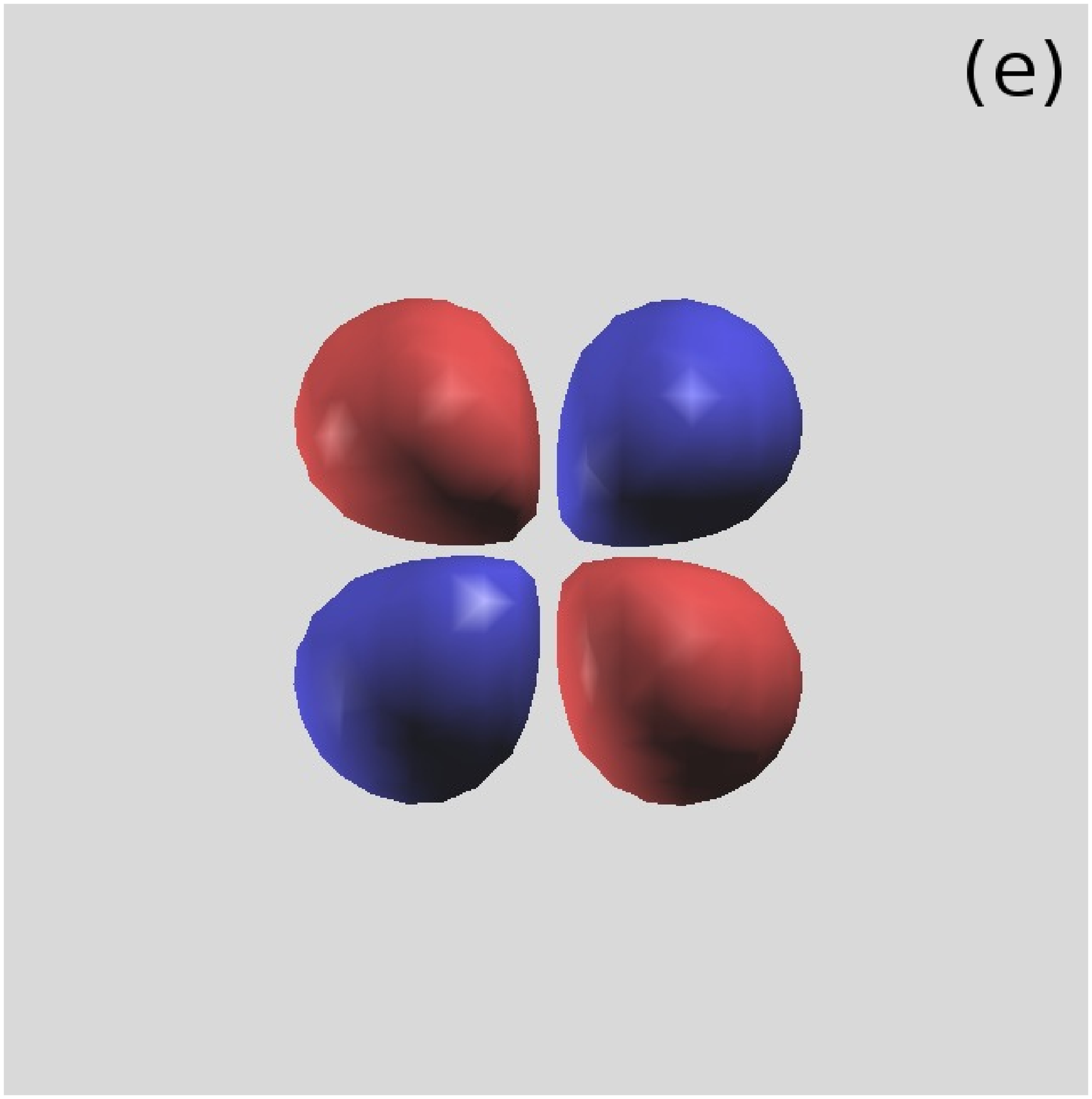}
\includegraphics[width=0.2\textwidth]{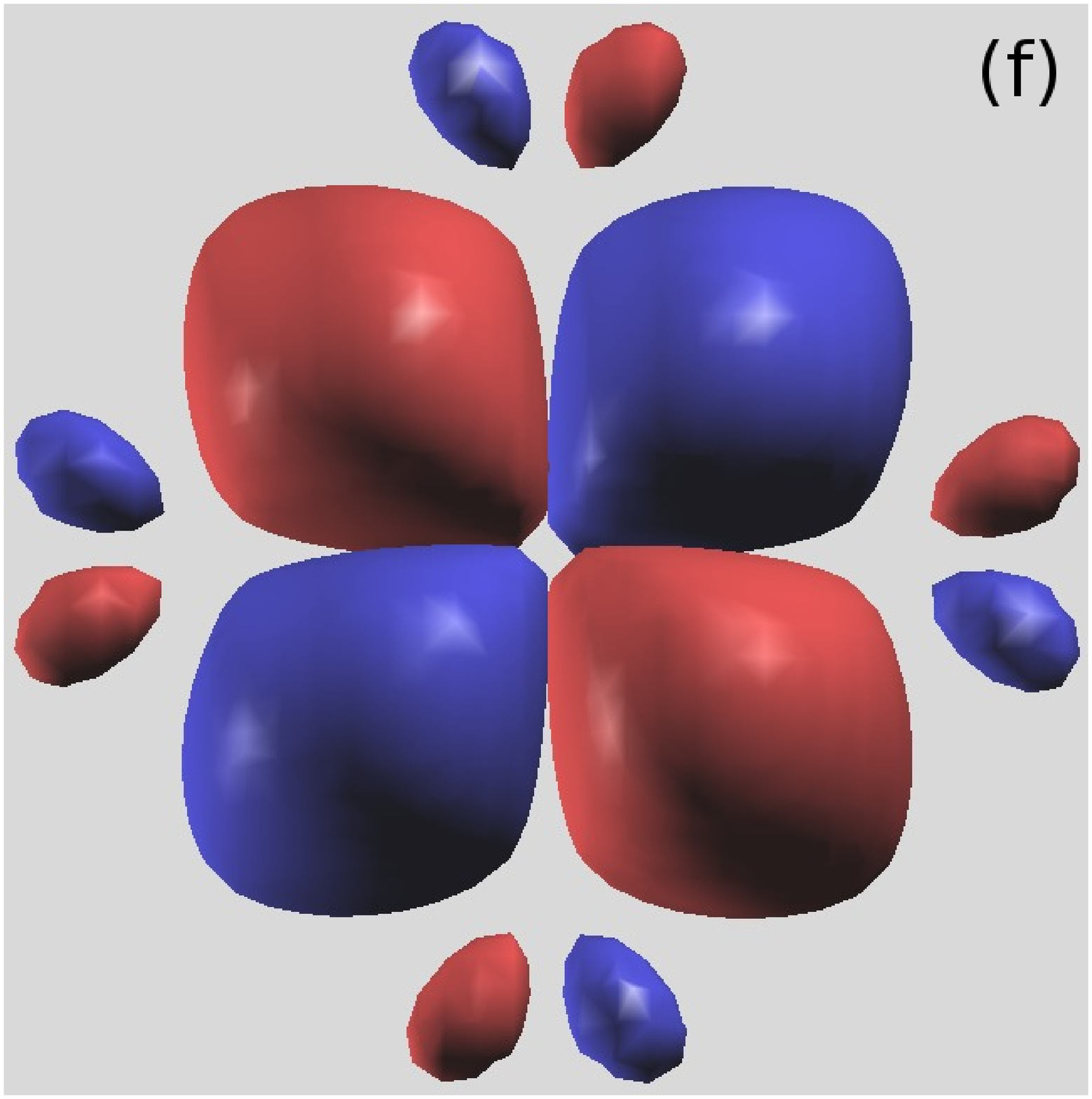}
\end{center}
\caption{(Color online) Contour-surface plot of the $d_{z^2}$ ((a),(b)), 
$d_{x^2-y^2}$ ((c),(d)) and $d_{xy}$ ((e),(f)) for different isovalues of
the MLWF. Figure prepared using XCRYSDEN \cite{xc}.}
\label{wf}
\end{figure}

This  {\it wannierization} \cite{wan1,wan2,wan3,wan4,wan5} procedure is 
implemented as follows. First, we select the group of Bloch bands from the 
spin unpolarized calculation for which the MLWF are calculated.  For
Fe$/$MgO, we take the valence bands as well as the 10 (counting spin) in-gap states.
These groups of bands do not overlap in energy with others, so that it is not 
necessary to perform the disentanglement procedure \cite{wan1,wan4}. 
Second, the  Bloch states $|\psi_{{\bf k}n}\rangle$ are projected over a 
set of localized functions. Based on the population analysis of the DFT 
calculation, we project over the atomic like $d$ orbitals centered around Fe 
and $p$ orbitals centered around oxygens.  In total, there are 96 $p$ orbitals 
(32 Oxygen atoms) and 5 $d$ orbitals. 
After an iterative procedure, the MLWF are determined.  Expectedly, the calculation yields  five  MLWF   localized around the Fe atom 
that,  in the neighborhood of the atom,  have the same symmetry that the real 
spherical harmonics with $\ell=2$, as shown in the left panels of 
Fig. \ref{wf} (a,c and e).  It is important to point out that the MLWF are 
not strictly identical to the atomic orbitals, because the tail of the wave 
functions have a different symmetry, as shown in the right panel  Fig. 
\ref{wf} (b,d and f). 

The representation of the DFT Hamiltonian in the basis of the MLWF yields a tight-binding Hamiltonian whose energy bands
are identical to the valence and in-gap bands obtained from DFT.  For the purpose of this work, we are interested in the intra-cell Hamiltonian:
\begin{eqnarray}
H_{DFT}=\left(\begin{array}{cc}
H_{dd} & V_{dp}\\
V_{pd} & H_{pp}
\end{array}
\right),
\end{eqnarray}
where $H_{dd}$ has dimension 5, corresponding to the $d$ orbitals of Fe,  and $H_{pp}$ has dimension 96, corresponding to the 3 $p$ orbitals of the 32 oxygen atoms in the unit cell. The $H_{dd}$  part describes the crystal field splitting of the $d$ levels. For the undistorted case, it describes the $t_{2g}$ triplet and $e_g$ doublet, separated by a crystal field splitting $\Delta_{CF}$.    Interestingly, diagonalization of $H_{dd}$ yields, for the undistorted case, 
$\Delta_{CF}=0.83$ eV, much smaller than the DFT splitting $1.45$ eV, that is only recovered if the whole $H_{DFT}$ matrix is diagonalized. Thus, we see that 
this approach permits us to quantify the ligand and crystal field contributions to the splitting. We find that 
almost half of the $t_{2g}-e_g$ splitting comes from the so called
 \cite{Abragam} ligand field contribution, described by $V_{dp}$, 
the hybridization between the $d$ like orbitals and the $p$ states that 
form the valence band of MgO. 

In order to preserve a small dimension of the Hilbert space,  so that the number of multi-electron configurations can be handled numerically, it is convenient to work with a truncated Hamiltonian for the $d$ electrons only,  but that includes their hybridization to the $p$ levels.  Such a Hamiltonian could be produced using degenerate second order perturbation theory for the different degenerate manifolds within the 5 $d$ levels, discussed below:
\begin{equation}
{\cal H}_{dd'}=  H_{dd'} +\sum_{p} \frac{V_{dp}V_{pd'}}{E_d-E_p}.
\end{equation}

This second order Hamiltonian yields eigenvalues within 10 percent of the exact ones. 
 It is possible to do better by realizing that the projection of the  exact eigenstates of $H_{DFT}$ over the  the $d$ like MLWF is always higher than 80 percent, and in most cases higher than 90 percent. 
More important, the spectrum and wave functions projected over the MLWF of the 5 in-gap states can be described with: 
\begin{equation}
H_{CF}=a(\ell_X^4 + \ell_Y^4 + \ell_z^4)  + d_2 l_z^2 + d_4 l_z^4,
\label{HCF1}
\end{equation}
\noindent where $l_{a}$ are the $\ell=2$ angular momentum matrices, and $d_2$, $d_4$  
and $a$ are obtained by fitting.
Here we approximate the MLWF by the real spherical harmonics centered in the Fe ion.
The same approximation is used in the calculation of spin-orbit and on-site Coulomb integrals later on. This methodology has been used before \cite{kk} with good qualitative results.

In order to fit $a$, $d_2$ and $d_4$ we employ the analytical expressions of the eigenvalues of $H_{CF}$:
 $18a+d_2+d_4, 18a+d_2+d_4, 18a+4d_2+16d_4, 24a, 24a+4d_2+16d_4$. 
  For the undistorted case,  the in-gap $d$ levels  obtained from diagonalization of Eq. (\ref{HCF1}) feature a triplet ($t_{2g}$) and a doublet ($e_g$), and 
  are fitted with   $d_2=d_4=0$, as expected from the octahedral symmetry.  The $t_{2g}-e_g$ splitting is thus given by $6a$, which yields $a=0.241$ eV.
  For the JT distorted case, the  $t_{2g}$ triplet is split into a singlet and a doublet, see Fig. \ref{fdosnm}a), while the $e_g$ doublet is also split. 
  The fitting yields  $a=0.250$ eV, $d_2=0.461$ eV and $d_4=-0.1$ eV. The difference between the fitted and computed energy levels are always smaller than 2 meV.

\section{Effective few electron Hamiltonian\label{CI}}

In the previous section  we have demonstrated that, starting from a DFT calculation for Fe in a supercell of MgO we 
are able to derive a crystal field Hamiltonian for the in-gap $d$ levels, including both crystal and ligand fields contributions, 
expressed in a basis of localized atomic-like orbitals provided by the 
maximally localized Wannier Functions. 

In this section we derive an effective spin Hamiltonian 
that accounts for the   low energy spectrum of a magnetic impurity within 
the MgO. This is done in two stages. We first build and solve a Hamiltonian 
for the  6 electrons in the $d$ levels or Fe, including the effect of crystal 
and ligand field as described at the DFT level, and adding the Coulomb and 
spin-orbit coupling interactions.  This few electron problem can be 
diagonalized numerically. In the second stage we analyze the symmetry and 
properties of the  low energy levels and propose an effective spin Hamiltonian 
that accounts for them. This is done for the undistorted and distorted 
configurations studied in Sec. II.   By so doing, we arrive at effective 
spin models in agreement with the literature \cite{Abragam,Ham77,Haupricht2010}.
 
\subsection{Multiplet calculation} 

We consider a Hamiltonian for the $N=6$ electrons in the $d$ orbitals of Fe in 
MgO that includes four terms, electron-electron, crystal-field 
and ligand field, spin-orbit and Zeeman interactions:

\beqa
H= H_{{\rm Coul}}+H_{\rm CF}+H_{\rm SO}+H_{\rm Zeem},
\label{Htot}
\eeqa

\noindent The Coulomb term reads:

\begin{equation}
H_{{\rm Coul}}=\frac{1}{2}\sum_{m,m'\atop n,n'}
V_{mnm'n'}
\sum_{\sigma\sigma'}d_{m\sigma}^\dag d_{n\sigma'}^\dag d_{n'\sigma'}d_{m'\sigma},
\label{hcoul}
\end{equation}
\noindent where $d_{m\sigma}^\dag$ ($d_{m\sigma}$) denotes the creation 
(annihilation) operator of an electron with spin $\sigma$  in the $\ell=2, 
\ell_z=m$ state of the magnetic atom, denoted by  $\phi_{m}(\vec{r})$, 
assumed to be equal to the product of a radial hydrogenic function (
with  effective charge $Z$ and a  effective Bohr radius $a_\mu$)
and a 
spherical harmonic. Thus, we are only considering $d^6$ configurations and leaving out
$pd^7 $ configurations. 

It turns out that the Coulomb integrals  $V_{mnm'n'}$ scale linearly with the value of $V_{0000}\equiv U$.  Explicit expressions for the  on-site Coulomb integrals are given in the appendix.  Specifically, $U$ could be computed using Eq. (\ref{scalingU}).  Another option, followed 
here, 
is to  consider $U$ as   an adjustable parameter.    In this work we  take $U=19.6$ eV,  which yields the correct
 splitting between the $^3P2$    excited state and the $^5D$ ground state of the free ion,  measured \cite{Sugar85} to be $2.41$ eV.

The second term in Eq. (\ref{Htot}) corresponds to the crystal and ligand fields 
Hamiltonian discussed in the previous section:
\beqa
H_{CF}=\frac{1}{2}\sum_{m,m',\sigma} \langle m|H_{\rm CF}|m'\rangle d_{m\sigma}^\dag d_{m'\sigma},
\label{hcf}
\eeqa
 with $ \langle m|H_{\rm CF}|m'\rangle $ derived from DFT using 
the procedure described above and, which is a very good approximation, is  given 
by Eq. (\ref{HCF1}).

The last term in the Hamiltonian describes spin-orbit coupling: 

\beqa
H_{\rm SO}=\zeta\,\sum_{mm',\sigma\sigma'} \langle m\sigma|\vec \ell\cdot \vec S|m'\sigma'\rangle
d_{m\sigma}^\dag d_{m'\sigma'},
\eeqa

\noindent where $\zeta$ is the single particle spin-orbit coupling of the 
$d$-electrons. It is also very frequently expressed as
$\lambda \vec L\cdot \vec S$ with $\vec L$ the total angular momentum. For the case of Fe$^{2+}$, both parameters $\zeta$ and $\lambda$ are related by $\lambda=-\zeta/2S$ \cite{Abragam},
with $\zeta=50.1$ meV and $S=2$.

The last term in Eq. (\ref{Htot}) corresponds to the  Zeeman Hamiltonian:
\beqa
\label{zeem}
&&\\ \nonumber
H_{\rm Zeem}=\mu_B \vec B\cdot \sum_{mm',\sigma\sigma'} \langle m,\sigma |\left(\vec l+g\vec S\right)
|m'\sigma'\rangle d_{m\sigma}^\dag d_{m'\sigma'},
\eeqa
\noindent where $g=2$. So, if we assume that the CF term is given by Eq. 
(\ref{HCF1}),  the multiplet Hamiltonian (\ref{Htot}) depends on five energy 
scales: $U$, $a$, $d_2,\;d_4 $ and $\zeta$ as well as the magnetic field.

For $N=6$ 
electrons, the total number of $d^6$ configurations is 210. Therefore, numerical 
diagonalization of the Hamiltonian is straightforward.
In agreement with Hund's rules, 
we obtain a ground state multiplet that maximizes $S$ and $L$. Thus, the 
ground state, denoted by $^5$D, has a degeneracy of $(2L+1)*(2S+1)=25$, with 
$L=S=2$. 
This low energy many body spectrum is fully independent of $U$ provided that the crystal field is not high enough to mix the $^5$D with the $^3$P2 multiplet. This could change if $d^7p$ configurations are included.

In order to analyze the results, it is convenient to add the different terms 
in the Hamiltonian one by one, in order of importance: Coulomb $U$, the 
crystal field ($a$, $d_2$,$d_4$), and spin-orbit coupling ($\zeta$).  Thus, in a 
first step  the problem is solved considering only $H_{\rm Coul}$. In this case the Hamiltonian commutes with $S^2$ 
and $L^2$, the square of total spin and orbital angular momentum.

\subsection{Undistorted F\MakeLowercase{e}/M\MakeLowercase{g}O}

\begin{figure}
  \begin{center}
\includegraphics[angle=270,width=1.\linewidth]{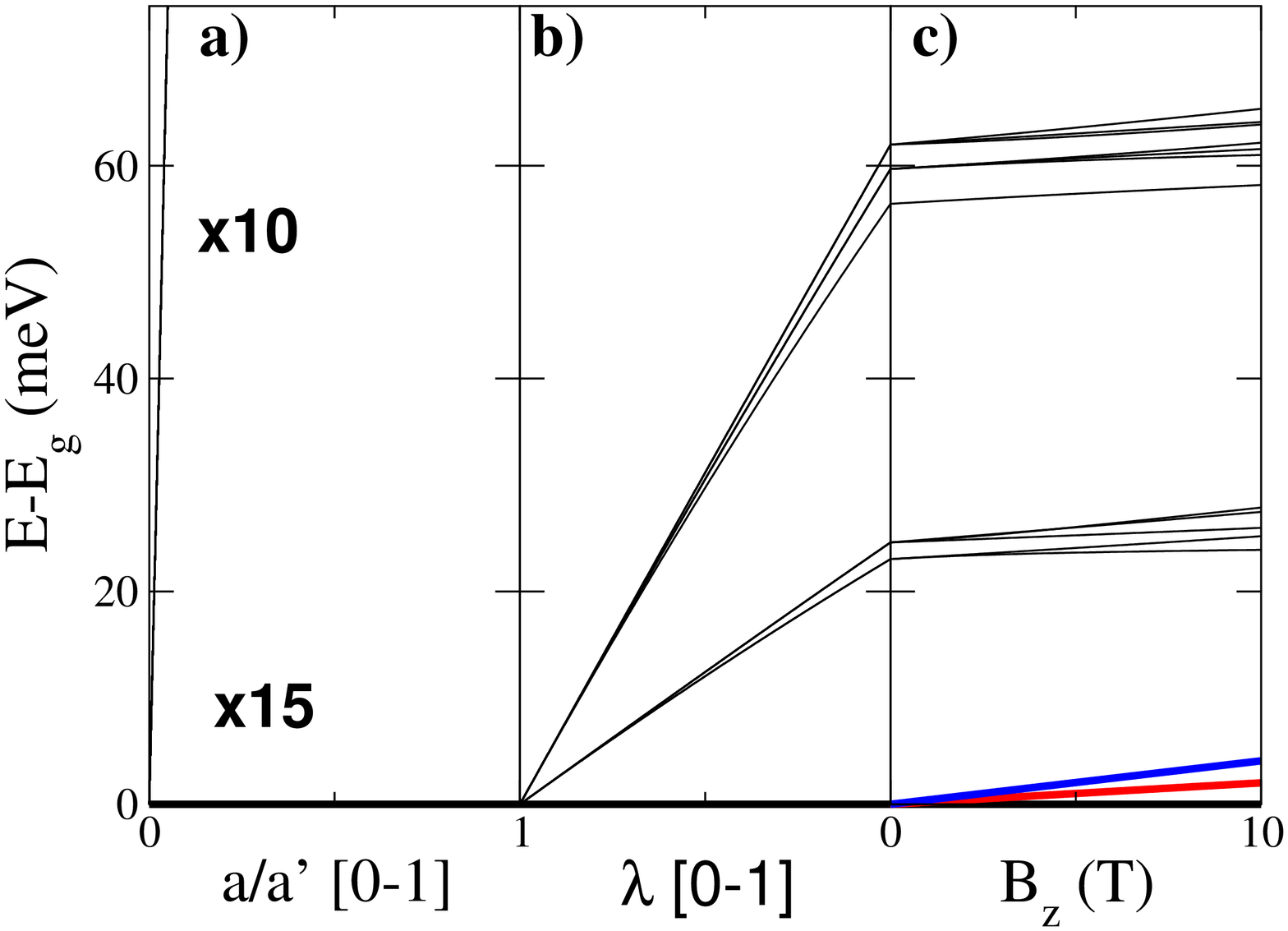}\\
\includegraphics[angle=270,width=0.8\linewidth]{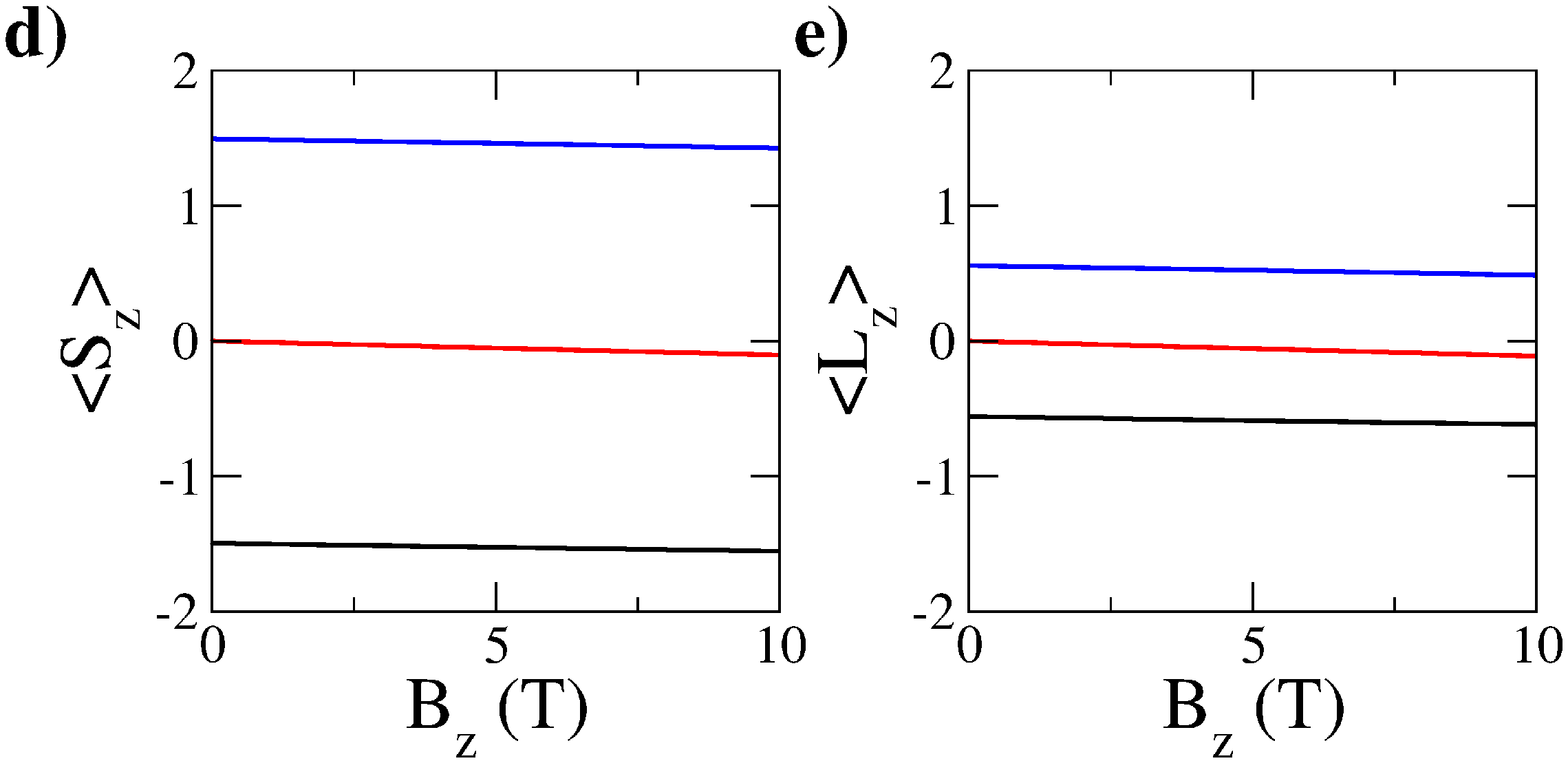}
  \end{center} 
  \caption{ {\bf a)} Energy splitting induced by the undistorted octahedral crystal field on the $(2S+1)(2L+1)=25$ times degenerate $^5D$ ground state versus the dimensionless cubic parameter $a/a'$ (with $a'=0.250$ meV). 
 {\bf b)}  and {\bf c)}  Energy splittings induced by the spin-orbit interaction and Zeeman  terms respectively. {\bf d)} Expectation value $\langle S_z\rangle$ and {\bf e)} $\langle L_z\rangle$ for the lowest three states for $a/a'=1$, $\zeta= 50$ meV.
In all cases, $U=19.2$ eV. 
  }
\label{fcisym}
\end{figure}

We discuss first the case of Fe$^{2+}$ in MgO without Jahn-Teller distortion.
The effect  of the octahedral component ($a$) of the crystal field on 
the $L=S=2$ multiplet is shown in Fig. \ref{fcisym}. As a result of the  
breaking of the orbital rotational symmetry, $L$ is no longer a good quantum 
number and the $2L+1$ degeneracy is partially lifted.  As we turn on $a$, see Eq. (\ref{HCF1}), 
the $^5$D levels of iron splits into two, an orbital $\Gamma_5$ triplet ground 
state, with total degeneracy 15,  and an orbital doublet excited state, 
10 times degenerate  see Fig.~\ref{fcisym}a). 

The  15-fold degeneracy of the ground state multiplet of the Fe in the octahedral environment of MgO
can be interpreted as if the ground state multiplet had  a 
$L=1$ orbital momentum. Actually, the 
representation of the $\vec{\ell}$ operator on the subspace of the $t_{2g}$ 
orbitals is isomorphic to the $\ell=1$ operators multiplied by $-1$ \cite{Abragam}.
When SOC is added to the Hamiltonian, the 15-fold degenerate ground state splits into a triplet, a quintuplet and a septuplet, in ascending energy order 
[see Fig. \ref{fcisym}b)].   This pattern can be rationalized in terms of the following effective Hamiltonian  where the total spin is coupled to the 
 pseudo angular momentum 
 ${\cal L}=-1$ \cite{Abragam}:
\begin{equation}
{\cal H}_{\rm eff1}= \lambda  
\vec{ \cal{L}}\cdot\vec{S} +\delta {\cal H}_{\rm eff1} ,
\label{LS}
\end{equation}
\noindent 
where $\lambda=-\zeta/2S$.
The first term naturally leads to a spectrum with multiplets
associated to $ \tilde J^2= (\vec{ \cal{L}}+\vec S)^2$: 
$\tilde J=1$ (ground), $\tilde J=2$ (first excited) and 
$\tilde J=3$ (third excited), with degeneracies $2\tilde J +1=3,5$ and $7$ 
respectively.  The result of the calculation of  the expectation values $\langle\psi|S_z|\psi\rangle$ for $\psi$ in the ground state triplet with $\tilde{J}=1$,  backs up the idea that  the $S=2$ spin is coupled to an  pseudo-angular momentum with ${\cal L}=-1$.  In Figs. \ref{fcisym}d) and e) we plot the expectation values of $S_z$ and $L_z$ for the 3 states of the ground state triplet as a function of the magnetic field. Notice that the $\tilde{J}_z=\pm 1$ and 
$\tilde{J}_z=0$ values are recovered by subtracting 
$\langle\psi|S_z|\psi\rangle$ and $\langle\psi|L_z|\psi\rangle$, 
in contrast with the common case of a total angular momentum.

The CI calculation for the 15 lowest energy states for  Fe$^{2+}$ in the undistorted environment  has some fine structure not captured by the first term 
in Eq. (\ref{LS}).  In particular, the multiplets with ${\tilde J}=1$ and $2$ 
have some fine structure [see Fig.\ref{fcisym}b)], that can be accounted 
by the a second term in the effective Hamiltonian
\begin{equation}
\delta {\cal H}_{\rm eff1} =
\tilde{a} (\tilde{J}_x^4 +\tilde{J}_y^4 +\tilde{J}_z^4) .
\label{self}
\end{equation}
\noindent This operator does not break the triple degeneracy of the ground 
state, breaks the $\tilde J=2$ into a triplet and a doublet (being 
isomorphic to the problem of the octahedral crystal field splitting of the 
$\ell=2$ orbitals),  and the $\tilde J=3$ into a singlet and two triplets.

In summary,  in the undistorted case, our calculation portraiys 
Fe$^{2+}$ as  a system with $S=2$ and pseudo-orbital 
momentum ${\cal L}=1$ \cite{Abragam}. Spin-orbit coupling leads to a ground state triplet  
with $\tilde J=1$.  The energy splitting to the first excited state, with 
$\tilde J=2$, is approximately linear in the atomic spin-orbit coupling, 
reflecting the fact that the octahedral symmetry quenches only in part the 
orbital angular momentum.  Thereby,  the effective model has to take into 
account $L$, and not only $S$.  The Jahn-Teller distortion, that we discuss 
next, changes this situation.

\subsection{Jahn-Teller distorted Fe/MgO}

\begin{figure}[t!]
  \begin{center}
\includegraphics[angle=270,width=1.\linewidth]{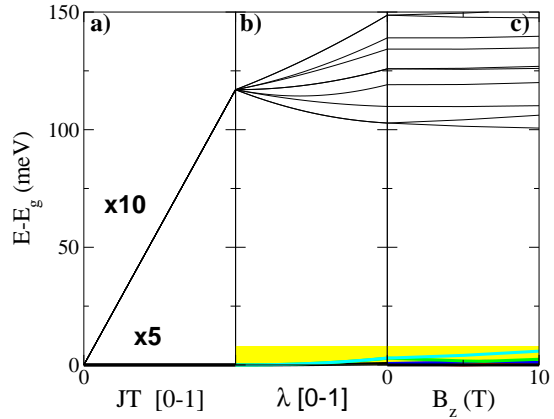}
  \end{center}
  \caption{{\bf a)} Energy splitting of the 15-fold degenerate orbital triplet in the octahedral crystal field induced by switching-on a deformation $d_2l_z^2+d_4l_z^4$.
 {\bf b)}  and {\bf c)}  Energy splittings induced by the spin-orbit interaction and Zeeman  terms respectively.
In all cases, $U=19.2$ eV, $\zeta= 50$  and $a=0.250$ eV. The five lower energy levels, corresponding to   $S=2$, 
appears in a yellow background.
    \label{figCI1}
  }
\end{figure}

\begin{figure}
  \begin{center}
  \includegraphics[width=0.75\linewidth,angle=-90]{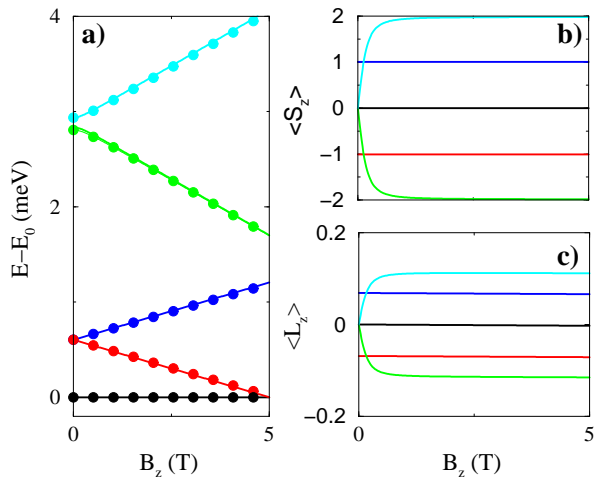} 
  \end{center}
  \caption{{\bf a)} (Solid lines) Low energy spectrum of the Fe$^{2+}$ ion in the MgO obtained using the DFT+WF Hamiltonian versus the magnetic field applied along the Jahn-Teller deformation axis ($z$).  The dots corresponds to the eigenvalues of the spin Hamiltonian in Eq. (\ref{seff}). Magnetic field dependence of the expectation values {\bf b)}, $\langle S_z\rangle$ and {\bf c)}, $\langle L_z\rangle$, for the five colored energy levels in {\bf a)}. In all cases, $U=19.2$ eV, $\zeta= 50$  and $a=0.250$ eV.
\label{figCI2}}
\end{figure}

We now discus the effect of the tetragonal distortion on the multiplet structure of Fe$^{2+}$ in MgO.  As discussed in Sec. \ref{sdft},   this distortion  introduces the uniaxial terms $d_2l_z^2+d_4l_z^4$ in Eq. (\ref{HCF1}).  The effect of the uniaxial terms on the many-body 15-fold degenerate ground state of the Hamiltonian  with $\zeta=0$ and $a=0.250$ meV is shown in Fig.~ \ref{figCI1}a). It is apparent that the JT distortion splits these 15 states into a ground state quintuplet, corresponding to a $S=2$ spin with quenched orbital momentum,  and a excited  manifold with 10 states.  Thus, it takes a JT distortion on top of the octahedral crystal field to eliminate the extra $2L+1=15$ degeneracy of the
$\Gamma_5$ orbital triplet. When  spin-orbit coupling is added  
[Fig. \ref{figCI1}b)] the $2S+1=5$ degeneracy is broken into a singlet, a doublet, and a split doublet [see Fig. \ref{figCI2}a)]. 
Finally, the Zeeman splitting breaks the remaining degeneracies, as observed in Fig. \ref{figCI1}c). 

Interestingly, the five low energy states, corresponding to $\tilde l_z=0$, can be described by an effective ${\cal S}=2$ Hamiltonian of the form
\beqa
{\cal H_{\rm eff2}}&=&D{\cal S}_z^2+\frac{a}{6}\left[
{\cal S}_x^4+{\cal S}_y^4+{\cal S}_z^4
\right]+g^* \mu_B \vec B\cdot \vec {\cal S}.
\label{seff}
\eeqa
 The comparison of the spectra as a function of a magnetic field $B_z$, calculated both with the full CI Hamiltonian and the effective spin model,    is shown in Fig. \ref{figCI2}a).  The parameters of the effective Hamiltonian are obtained by fitting
  to the multiplet calculation. We obtain 
  $D=0.734$ meV, $a=0.130$ meV and $g^*=2.03$. 
The expectation values of $S_z$ and $L_z$, computed with the eigenstates of the full CI Hamiltonian,  are shown in  Fig. \ref{figCI2}b),c) as a function of $B_z$.  It is apparent that the ground state (black line) has $S_z=0$, as a result of  
the dominant uniaxial term $D{\cal S}_z^2$ favoring the minimum spin projection as ground state, ${\cal S}_z=0$.
The first excited doublet, split by $B_z$, has $S_z=\pm 1$. The ${\cal S}_x^4 +{\cal S}_y^4$ term  couples the otherwise degenerate doublet $S_z=\pm 2$, resulting in a quantum spin tunneling splitting.  The mixing of the wave functions is apparent in the non-linear evolution of the expectation value of $\langle S_z\rangle$ as a function of $B_z$.  At small field the magnetic moment is quenched. At higher field the Zeeman contribution overcomes the quantum spin tunneling.  We note in passing that, in contrast with  the  $S=2$  spin  with $C_2$ in plane symmetry \cite{Sessoli},    in our case there is no quantum spin tunneling splitting within the $S_z=\pm 1$ doublet, that remains degenerate.

\section{Discussion and Conclusions \label{con}}

The results of the previous sections illustrate how, for the cases of Fe$^{2+}$ in MgO with and without Jahn Teller distortion, we have been able to derive effective spin Hamiltonians [Eq. (\ref{LS})  and Eq. (\ref{seff})] that reproduce the spectra obtained from the few-electron Hamiltonian. The parameters are derived directly from a DFT calculation of the electronic structure of this system. 
We now list possible improvements for the method. In addition, we 
briefly discuss the implications for a technologically relevant system, MgO tunnel barriers with Fe electrodes \cite{Butler_Zhang_prb_2001,Mathon_Umerski_prb_2001,yua04,par04}, and present our conclusions.

\subsection{Improvements for the method}

There are several ways in which the method presented in this manuscript could be improved. First, the approximation that the Wannier functions are atomic orbitals in the evaluation of the matrix elements of both the spin-orbit coupling and Coulomb interaction could be avoided at the price of performing the  numeric integration using  the actual Wannier functions.  This would also allow extending the method to situations in which the localized atomic orbital lives in interstitial sites, such as the technologically relevant \cite{NV1,NV2} cases of NV centers in diamond \cite{Louie2012}, or Mg vacancies in the MgO \cite{Araujo_Kapilashrami_apl_2010,Wang_Pang_prb_2009}.  Second, a more accurate quantitative description would require us to correct the double counting of some of the Coulomb interactions \cite{Lichtenstein_Katsnelson_prb_1998,cf-wan,davidjacob,kulik06}.
Third,  the Hilbert space in the multiplet calculation could be expanded in two ways, either including more intra-atomic
configurations \cite{Heinrich2014} , such as   $pd^5$,  as well as configurations where the charge is transferred to the neighbor oxygen atoms \cite{Haverkort12,Fernandez2014}.   Fourth, the GGA calculation underestimates the gap of insulators, which  most likely has some influence on the $d$ levels as well.  The use of a hybrid functional, or of an approximation adequate to compute energy gaps, such as the GW method \cite{hyb,hyb1,gw,refe3,refe4}  would be an improvement, but the computational overhead for unit cells with tens of atoms is far from small.   
Finally, the method presented here could be improved obtaining $U$ 
from first principles calculations \cite{refe1,refe3,refe4,refe5}.

\subsection{Influence of F\MakeLowercase{e} impurities in the barrier M\MakeLowercase{g}O of a magnetic tunnel junction\label{sTMR}}

We now briefly discuss some relevant consequences drawn from our calculation in the context of spin dependent transport in MgO magnetic tunnel junctions  with Fe based electrodes such  as Fe or CoFeB \cite{yua04,par04}.
A key figure of merit of magnetic tunnel junctions is the magnetoresistance, 
 defined as $MR=100\times (R_{AP}-R_{P})/R_{P}$, where $R_{P}$ and $R_{AP}$ are the resistance for parallel and antiparallel orientation of the electrode magnetizations.   A very large MR, exceeding 1000, was predicted for Fe/MgO/Fe  MTJ \cite{Butler_Zhang_prb_2001,Mathon_Umerski_prb_2001}.  Actual experiments in this system have found room temperature MR above 600 \cite{Ikeda_Hayakawa_apl_2008}. that have permitted a tremendous boost of this technology, but  quite below the theoretical limit.
 
  The very likely  presence of substitutional impurities of Fe in the MgO barrier would affect transport in two ways, opening
    two additional  tunneling channels in the  magnetic tunnel junction. On one side, electrons could tunnel  through the  in-gap $d$ levels 
(see Fig. \ref{fdosp}).
 Elastic tunneling through these states is possible  at large bias ($\simeq 1eV$),  when the Fermi energy of one of the electrodes is set on resonance with the in-gap $d$ levels.  This would yield characteristic resonance  line shapes at finite bias, not much different from those observed experimentally \cite{Timopheev_Pogorelov_prb_2014}.  At small bias, electrons could still tunnel through these $d$ levels through second order cotunneling processes, in which the transport electron would excite a spin transition  between the low energy states of the Fe, within a range of a few meV, see Fig. \ref{figCI2} (a) . Whereas this process will give a much smaller contribution to transport, they are known to be an efficient \cite{Delgado_Rossier_prb_2011}  source of spin-flip.  These problems will be addressed qualitatively elsewhere.  

\subsection{Summary}

In summary we have presented a method to derive effective spin Hamiltonians for magnetic atoms inside insulators, starting from a DFT calculation based on plane waves.  This is achieved by post-processing the DFT calculation to obtain the maximally localized Wannier functions, which, in the system considered here, happen to be atomic-like orbitals in the magnetic atom.  Expressed in the basis of the Wannier functions, we can build a many-body Hamiltonian [Eq. (\ref{Htot})] that includes the effect of crystal and ligand fields, as given by DFT, and the effect of spin-orbit interaction and on-site Coulomb repulsion at the magnetic atom.  This model is solved by numerical diagonalization. An analysis of the symmetry of the spectrum and the multi-electron wave functions allows us to postulate a much simpler  effective spin Hamiltonian [Eq. (\ref{LS})  and Eq. (\ref{seff})]  that accurately describe the low energy sector of the spectrum.  We apply this  method to the case of Fe$^{2+}$ in MgO, considering both the undistorted and distorted geometries. In the former the orbital momentum is not quenched which results in very different type of effective Hamiltonian, featuring both $S$ and $L$ operators. In the Jahn Teller distorted case, orbital momentum is quenched, and a spin $S=2$ Hamiltonian is enough to describe the lowest energy states of 
Fe$^{2+}$.  The method can be implemented to study a variety of systems, including diluted magnetic semiconductors, magnetic adsorbates on insulating surfaces, and  magnetic atoms migrated from the electrodes into the barrier in magnetic tunnel junctions.

\section{Acknowledgements}
We acknowledge J. W. Gonz\'alez for technical assistance with the use of Quantum Espresso and Wannier90.  We also acknowledge J. L. Lado for  fruitful discussions and assistance with technical aspects of DFT. 
AF acknowledges funding from the European Union's Seventh Framework 
Programme for research, technological development and demonstration,  
under the PEOPLE programme, Marie Curie COFUND Actions, grant agreement number 
600375 and CONICET. JFR acknowledges  financial support by  Generalitat Valenciana 
(ACOMP/2010/070), Prometeo.

\appendix
\section{Evaluation of the Coulomb integrals}
\label{aint}
The Coulomb parameters $V_{ijkl}$ are calculated assuming n$d$ hydrogen-like wavefunctions
\beqa
\\ \nonumber
V_{ijkl}=\frac{e^2}{4\pi \epsilon_0}\int d\vec r_1 d\vec r_2\frac{ \phi_i^*(\vec r_1) \phi_j^*(\vec r_2) \phi_k(\vec r_2)\phi_l(\vec r_1)}{\left|\vec r_1-\vec r_2\right|},
\label{vijkl}
\eeqa
where
\beqa
\phi_i(\vec r)=R_{n,2}(r) Y_2^i(\Omega),
\eeqa
with $Y_l^m(\Omega)$ the spherical harmonic and $R_{n,2}(r)$ the hydrogen wavefunction corresponding to quantum numbers $n$ and $l=2$ for an effective nuclear charge $Z$ and effective radius $a_\mu$
\beqa
\label{rad}
R_{n2}(r)=\frac{4}{27\sqrt{10}}\left(z/3\right)^{3/2}\left(zr\right)^2e^{-zr/3}
\eeqa
where $z=Z/a_\mu$. Using the spherical harmonic expansion
\beeq
\frac{1}{\left|\vec r_1-\vec r_2\right|}=4\pi \sum_{l=0}^\infty \sum_{m=-l}^{+l}\frac{1}{2l+1}\frac{r_<^l}{r_>^{l+1}}Y_l^{m*}(\Omega_1)Y_l^m(\Omega_2),
\eeqe
where $r_< ={\rm Min}(r_1,r_2)$ and  $r_> ={\rm Max}(r_1,r_2)$ ,
one can divide the integral in Eq. (\ref{vijkl}) into an angular part and a radial part, writing then
\beeq
V_{ijkl}=\sum_{\ell} U_{\ell} \chi_{ijkl}^{\ell,m},
\eeqe
where $U_{\ell}$ and $\chi_{ijkl}^{\ell,m}$ contains the radial and the angular information respectively.
 The angular integrations over the solid angles $\Omega_1$ and $\Omega_2$ factorizes, $\chi_{ijkl}^{\ell,m}=(-1)^{i+j+m} \Phi_{2,\ell ,2}^{-i,m,l}  \Phi_{2,\ell ,2}^{-j,-m,k}$, where each part can be written in terms of the Wigner 3-j symbols \cite{Wigner_book_1959}
\beqa
\Phi_{l_1,l_2,l_3}^{m_1,m_2,m_3}=&&\\ \nonumber
\int d\Omega && Y_{l_1}^{m_1}(\Omega) Y_{l_2}^{m_2}(\Omega)Y_{l_3}^{m_3}(\Omega)=
\crcr
&&
\sqrt{\frac{(2l_1+1)(2l_2+1)(2l_3+1)}{4\pi}}\left(
\begin{array}{ccc}
 l_1 & l_2 & l_3 \\
0 & 0 & 0 
\end{array}
\right)
\crcr
&&\times
\left(
\begin{array}{ccc}
 l_1 & l_2 & l_3 \\
m_1 & m_2 & m_3 
\end{array}
\right).
\eeqa
The radial part is given by
\beqa
U_{\ell}\equiv\frac{e^2}{\epsilon_0(2\ell +1)}\int dr dr' r^2r'^2\frac{r_<^\ell}{r_>^{\ell+1}} R_{n2}^2(r) R_{n2}^2(r').
\crcr
\label{scalingU}
\eeqa
This integral is solved numerically for $l=0,2,$ and $4$. From Eqs. (\ref{scalingU}) and (\ref{rad}) it is clear that all matrix elements $V_{ijkl}$ scales proportional to $z$. For convenience, instead of using $z$ as a free parameter, we use $U=V_{0000}$ as the free parameter. In particular, $z=1.95/a_0$, with $a_0$ the Bohr radius, for $U=19.6$ eV.

\end{document}